# Thermal Transport Properties of Nanoporous Silicon with Significant Specific Surface Area

Mykola Isaiev[1c], Yuliia Mankovska[2*], Vasyl Kuryliuk[2], and David Lacroix[1]

[1]Université de Lorraine, CNRS, LEMTA, Nancy F-54000, France

[2]Faculty of Physics, Taras Shevchenko National University of Kyiv, 64/13, Volodymrska Str., 01601 Kyiv, Ukraine

*Current affiliation: Laboratoire d'Optique Appliquée – ENSTA Paris, Ecole Polytechnique, CNRS, IP Paris, Palaiseau, France

cCorresponding author: mykola.isaiev@univ-lorraine.fr

This paper studies thermal transport in nanoporous silicon with a significant specific surface area. First, the equilibrium molecular dynamics approach was used to obtain the dependence of thermal conductivity on a specific surface area. Then, a modified phonon transport kinetic theory-based approach was developed to analyze thermal conductivity. Two models were used to evaluate the phonon mean free path in the porous materials. The first model approximates the dependence of the mean free path only with the specific surface area, and the second one investigates the dependence of the mean free path variation with the porosity in the peculiar case of a highly porous matrix. Both models approximate molecular dynamics data well for the smaller porosity values, while the first model fails for large porosities. The second model matches well with molecular dynamics simulations for all considered ranges of the porosities. This work illustrates that the phonon mean free path dependence with the porosity/volume fraction of composite materials is essential for describing thermal transport in systems with significant surface-to-volume fractions.

The miniaturization of electronic devices and their components nowadays leads to issues such as overheating with hotspots and subsequent failure occurrence. Therefore, further development of such technologies requires significant efforts to establish the background of efficient thermal management in nanostructured objects. The latter involves understanding heat carriers transfer mechanisms, scattering and interactions close to an interface separating different species. Porous silicon (PS) is an excellent candidate to be a model object to investigate various conditions favoring/disfavoring thermal transport [1,2]. From a practical point of view, one can fabricate PS with a wide range of pore sizes, porosity, and morphology[3]. Thus, it gives an excellent and clear basis for experimental verifications of the developed models[4,5]. Further, such results can be broadcasted for various nanocomposite systems, where interfacial also crucially impacts thermal transport[6,7].

Specifically, the significant reduction of thermal conductivity of PS compared to the bulk one is already well known[8,9]. Two main reasons explain this: i) the natural removal of material and ii) the phonon scattering at the interfaces. While the first mechanism is dominant for large pore radius, the second is more crucial while decreasing the pore radius and thus increasing the phonon scattering specific surface area.

One of the models that describe this tendency is Minnich and Chen model [10]; in this model, the effective thermal conductivity of a nanoporous system ($\lambda$) can be decomposed into multiplicands:

$$\lambda = \lambda_{nm} F, \qquad (1)$$

where $\lambda_{nm}$ is the thermal conductivity of the nanoporous matrix, which differs from the bulk one due to the phonons mean free path reduction as a result of the scattering on the interfacial boundaries; $F$ is the factor describing the reduction of thermal conductivity due to the decrease the portion of the bulk material. According to the Maxwell model for effective media, this factor can be presented as follows for porous material [11]:

$$F = \frac{2 - 2P}{2 + P}, \qquad (2)$$

where $P$ is the porosity.

Regarding the thermal conductivity of the nanoporous matrix, Minich and Chen proposed a simple approach for the evaluation of the thermal conductivity of the matrix as follows:

$$\lambda_{nm} = \frac{1}{3} C_h v_h \frac{1}{1/\Lambda_h + \Phi/4}, \qquad (3)$$

where $C_h$ is the volumetric heat capacity of the host matrix, $v_h$ is the mean velocity of phonons of the host matrix, $\Lambda_h$ is the mean free pass in the host matrix material, and $\Phi = \frac{A_{por}}{V_{box}} = 4\pi R^2/L^3$ is the specific surface area of the pore. Eq. (3) is based on several assumptions, and one of them is that the cross-section of the phonons' scattering at the pore edge has the following form:

$$\sigma = \pi R^2, \qquad (4)$$

where $R$ is the pore radius. This cross-section corresponds to the assumption of the phonon scattering at the edge of a solid sphere. Yet, this assumption forms the bottleneck of the approach and its application to the cases of small pore radii with a very high interface density, where the multiple scattering events are frequent.

Therefore, our work aims to understand the role of multi-scattering processes in the perturbation of heat transfer in nanoporous materials. We first simulated thermal conductivity in porous silicon using the equilibrium molecular dynamics (EMD) approach. Then, we adapted the kinetic theory (KT) approach to model thermal conductivity in PS, considering the contribution of each mode to heat transfer. The latter allows us to identify the role of different mechanisms leading to the modification of the thermal transport performance of the porous matrix.

In this work, we investigated a crystalline silicon matrix with a lattice constant equal to $a = 5.431$ Å for molecular dynamics simulations; the simulation domains contain a repeated translation of silicon cells in x, y, and z directions. We chose the number of repetitions ($n$) to equal 6, 8, 10, and 12, and the domain size was $L = n\,a$. Periodic boundary conditions were set in all directions. Pore structure was created by cutting out the atoms in a sphere with a radius $R$ located in the center of the simulated domain. The sphere radius was considered in the range from 0 (bulk silicon) to $(L-a)/2$ with the step $a/2$. Such variation of $L$ and $R$ allows us to consider porosity and specific area of nanoporous material in a wide range. The interaction between silicon atoms was simulated with the respect of Tersoff potential[12].

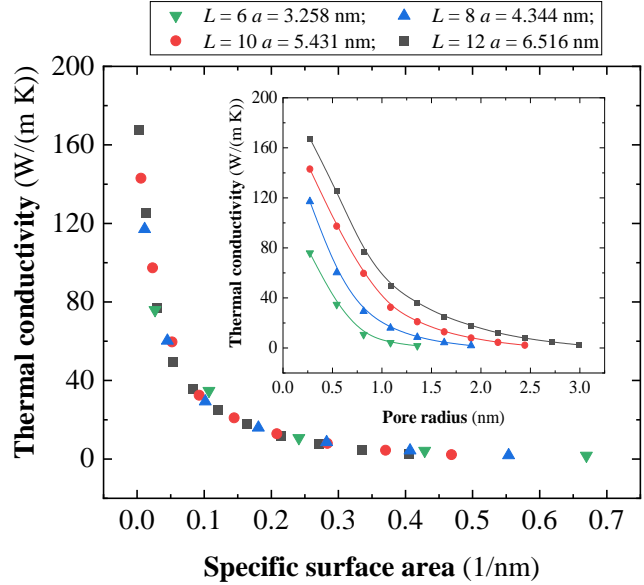

**Fig. 1.** Dependence of thermal conductivity with the specific surface ratio of a pore. Inset: Dependence of thermal conductivity with pore radius

We used the Green-Kubo formalism for the thermal conductivity calculations:

$$\lambda = \frac{1}{3Vk_BT}\int_0^{t_c} dt \langle \bm{J}(0)\bm{J}(t)\rangle_{t_s}, \qquad (5)$$

where $V$ is the volume of the simulation domain, $T$ is the temperature, $\bm{J}(t)$ is the heat flux vector, $t_c$ is the finite correlation time for which integration was carried out, $t_s$ is the sampling time for sampling time over which the autocorrelation function was accumulated for averaging.

Fig. 1 presents the dependence of thermal conductivity with the pore specific surface area for different box sizes. The inset of Fig. 1 details the dependence of the thermal conductivity as a function of the pore radius. As shown in Fig. 1, the specific surface ratio is the more versatile parameter for describing thermal conductivity reduction in the considered pore radius range.

In order to evaluate the Minich and Chen model [10] for thermal transport in composite media, we used the kinetic theory (KT) approach developed previously by P. Chantrenne et al[13]. In the frames of this model, the thermal conductivity of the nanoporous matrix can be represented as follows:

$$\lambda_{nm} = 6\frac{k_B}{3\frac{a^3}{4}}\int_0^\infty d\omega C_V(\omega)\, D(\omega) \sum_p \xi(\omega,p)v^2(\omega,p)\tau(\omega,p), \qquad (6)$$

https://doi.org/10.48550/arXiv.2302.13847

where $k_B$ is the Boltzmann constant, $\omega$ and $p$ are the phonon frequency and the polarization, $C_V(\omega)$ is the per mode per unit volume heat capacity, $D(\omega)$ is the phonons density of states, $v(\omega, p)$ is the phonon group velocity, $\tau(\omega, p)$ is the phonon relaxation time due to phonons scattering, $\xi(\omega, p)$ is the coefficient to express the contributions of different polarizations to the phonons density of states.

Considering the classical nature of MD computations, we use the KT model with constant (high-temperature limit) heat capacity to minimize the number of different factors in KT and MD approaches. $D(\omega)$, $v(\omega, p)$ and $\xi(\omega, p)$ were taken from our previous paper [14]. According to the Matthiessen rule, the resulting relaxation time (RT) was calculated as follows

$$\tau^{-1}(\omega, p) = \tau_{ph-ph}^{-1} + \tau_{por}^{-1}, \qquad (7)$$

where $\tau_{ph-ph}$ is the lifetime due to phonon-phonon scattering; this dependence was also taken from [14]; $\tau_{por}$ is the lifetime due to phonon scattering at the pore edge. In our calculations, we evaluate this lifetime as follows

$$\tau_{por} = \frac{l_{por}}{v(\omega, p)}, \qquad (8)$$

where $l_{por}$ is the mean free path due to phonon scattering at the pore edge. Following Minich and Chen [10], it was estimated as

$$l_{por} = l_{Minnich} = \frac{4}{\Phi} = \frac{L^3}{\pi R^2}. \qquad (9)$$

Fig. 2 presents the dependence of the thermal conductivity of the nanoporous matrix calculated with the KT approach by Eq. (6) and the one evaluated from MD simulations by Eq. (1).

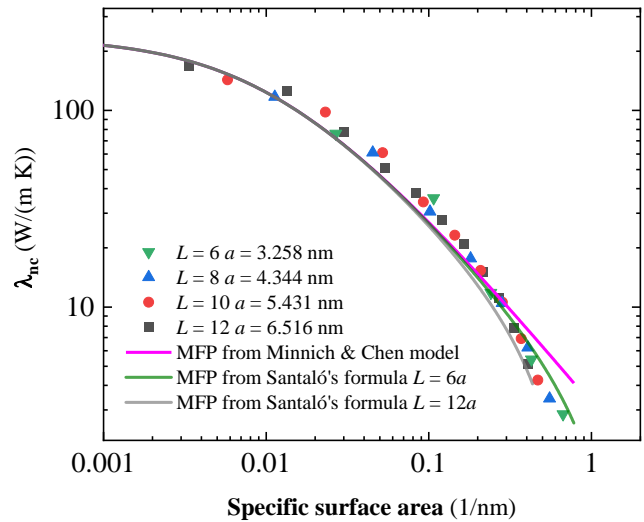

**Fig. 2.** The dependence of the thermal conductivity of the nanoporous matrix calculated with the KT and MD approaches

As shown in Fig. 2, the Minich and Chen model approximates well MD data for the small and intermediate regions of specific surface area (before 0.3 nm$^{-1}$). While for the high interfacial density, the difference in thermal conductivity may reach 100%. Therefore, one can conclude that phonon localization may play a significant role in high interface densities. Such localization may arise as a result of decreasing distance between pores due to the necking effect[15,16]. The distance decreases mainly because of the increased volume occupied by a pore.

In order to take this into account, we use the approximation of the mean free path obtained for the periodic Lorentz gas represented by two types of particles. The first type is the "heavy particles" periodically located in a 3d lattice representing the pores, while the "light particles" represent the phonons. For such a configuration, the mean free path of the light particle due to the scattering on heavy one can be represented by Santaló's formula [17,18]:

$$l_{Santaló} = \frac{L^3 - \frac{4}{3}\pi R^3}{\pi R^2} = (1-P)l_{Minnich}, \qquad (10)$$

where $P$ is the porosity. As one can see, Santaló's formula for the mean free path also shows the impact of reducing the material's volume on the mean free path. This impact is crucial for highly porous materials with an enormous specific surface area. One should note that Eq. (10) was obtained for specular scattering of the light particle at the surface of the heavy one (billiard model).

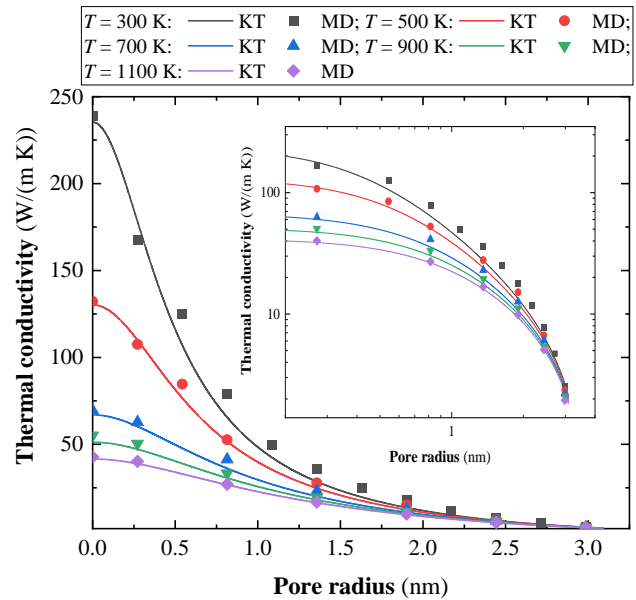

**Fig. 4** Dependence of thermal conductivity with pore radius for different temperatures (the case $L = 12\,a$)

The analytical description of the modified mean free path with Santaló's formula is also presented in Fig. 2. As one can see, this model well describes the dependence of thermal conductivity for higher values of the specific surface area. It should be noted that in Fig. 2, we presented the curve only for two extreme cases $L = 6a$ and $L = 12a$. As shown in Fig. 2, the impact of the volume reduction is more pronounced for the bigger box for the same specific surface area. However, the deviation of both curves from the one given by the Minnich and Chen approximation is significant only under the enormous specific surface area, which may arise only at the nanoscale.

The KT model with the mean free path modification based on the "billiard model" also describes the temperature dependence of porous silicon (see Fig. 4).

In conclusion, we performed MD and KT simulations of thermal conductivity in porous silicon for different specific surface areas. We found a significant reduction of TC for large specific surface ratio. In the latter case the prediction based on the Minnich and Chen model fails. Thus, Santaló's formula used to evaluate the mean free path in a Lorentz gas was adopted to approximate the mean free path in high porosity silicon. This formula considers the dependence of the mean free path with the porosity istself. Specifically, such dependence arise due to the necking effect – decreasing the distance between pores. The analytical KT model with the mean free path estimation by the Santaló's formula predicts well the variations of thermal conductivity for higher porosity. Thus, we can state the crucial impact of the volume fraction of the composite components at the nanoscale on the phonon mean-free reduction. The latter should be considered for thermal engineering in nanoelectronics, where the components' size is currently trending to several nanometers[19].

**Acknowledgment.**

This paper contains the results obtained in the frames of the projects ''Hotline'' ANR-19-CE09-0003 and "FASTE" ANR-22-CE50-002. This work was performed using HPC resources from GENCI-TGCC and GENCI-IDRIS (A0130913052), in addition HPC resources were partially provided by the EXPLOR center hosted by the Université de Lorraine. Thanks to ''STOCK NRJ'' that is co-financed by the European Union within the framework of the Program FEDER-FSE Lorraine and Massif des Vosges 2014–2020.